\documentclass[conference]{IEEEtran}
\IEEEoverridecommandlockouts
\usepackage{cite}
\usepackage{amsmath,amssymb,amsfonts}
\usepackage{algorithmic}
\usepackage{graphicx}
\usepackage{textcomp}
\usepackage{hyperref}
\usepackage{float}
\usepackage{booktabs}
\usepackage{array}
\usepackage{soul}

\def\BibTeX{{\rm B\kern-.05em{\sc i\kern-.025em b}\kern-.08em
    T\kern-.1667em\lower.7ex\hbox{E}\kern-.125emX}}

\usepackage[dvipsnames]{xcolor} 

\usepackage{listings}
\usepackage{adjustbox}
\usepackage{tcolorbox}  
\usepackage{mdframed}   

\renewcommand{\figureautorefname}{Fig.}
\renewcommand{\thefigure}{\arabic{figure}}

\definecolor{codeadd}{rgb}{0.9, 1.0, 0.9}
\definecolor{coderemove}{rgb}{1.0, 0.9, 0.9}

\lstdefinestyle{pythonstyle}{
    language=Python,
    basicstyle=\scriptsize\ttfamily,
    numbers=left,
    numberstyle=\tiny,
    numbersep=5pt,
    xleftmargin=2em,
    breaklines=true,
    postbreak=\mbox{\textcolor{red}{$\hookrightarrow$}\space},
    showstringspaces=false,
    escapeinside={(*@}{@*)}, % for custom highlights
}

\newcommand{\datasetname}{{\textit{PyResBugs}}}
\newcommand{\datasetnumberofbugs}{{$5,007$}}

\usepackage{tikz}
\usepackage{scalerel}

\begin{document}

\raggedbottom

%\title{\datasetname{}: A Dataset of Residual Bugs with Natural Language Annotations for Fault Injection}

\title{\datasetname{}: A Dataset of Residual Python Bugs for Natural Language-Driven Fault Injection}
%\title{Exploring Natural Language-Driven Fault Injection with \datasetname{}: A Python Residual Bugs Dataset}

% Natural Language-Driven Fault Injection with \datasetname{}: A Dataset of Python Residual Bugs

\author{\IEEEauthorblockN{Domenico Cotroneo, Giuseppe De Rosa, Pietro Liguori}
\IEEEauthorblockA{\textit{University of Naples Federico II}\\
Naples, Italy \\
cotroneo@unina.it, giuseppe.derosa20@unina.it, pietro.liguori@unina.it}}

\begin{comment}
\author{\IEEEauthorblockN{Domenico Cotroneo}
\IEEEauthorblockA{\textit{DIETI} \\
\textit{University of Naples Federico II}\\
Naples, Italy \\
cotroneo@unina.it}
\and
\IEEEauthorblockN{Giuseppe De Rosa}
\IEEEauthorblockA{\textit{DIETI} \\
\textit{University of Naples Federico II}\\
Naples, Italy \\
giuseppe.derosa20@unina.it}
\and
\IEEEauthorblockN{Pietro Liguori}
\IEEEauthorblockA{\textit{DIETI} \\
\textit{University of Naples Federico II}\\
Naples, Italy \\
pietro.liguori@unina.it}
}
\end{comment}

\maketitle
\thispagestyle{plain}
\pagestyle{plain}

\begin{abstract}
This paper presents \datasetname{}, a curated dataset of \textit{residual} bugs, i.e., defects that persist undetected during traditional testing but later surface in production—collected from major Python frameworks. Each bug in the dataset is paired with its corresponding fault-free (fixed) version and annotated with multi-level natural language (NL) descriptions. These NL descriptions enable natural language-driven fault injection, offering a novel approach to simulating real-world faults in software systems. 
By bridging the gap between Software Fault Injection techniques and real-world representativeness, \datasetname{} provides researchers with a high-quality resource for advancing AI-driven automated testing in Python systems.
\end{abstract}

\begin{IEEEkeywords}
Residual Bugs, Dataset, Python, Fault Injection, Natural Language 
\end{IEEEkeywords}

\section{Introduction}
\label{sec:intro}
% Context: SFI
Modern society's dependence on software systems, powering critical services in healthcare, finance, and transportation, has grown alongside the increasing complexity of software. This complexity makes defect-free systems unattainable, as demonstrated by high-profile failures disrupting economies and services~\cite{businessinsider2021citi, cnn2018uber, cnn2024crowdstrike,ieee2019boeing737, dolfing2019knightcapital}. While rigorous testing mitigates many issues, fault-tolerance mechanisms are essential to handle the unforeseen faults that inevitably arise in complex systems~\cite{nasa2004software, iso2011productdevelopment}, especially in weakly typed or interpreted languages like Python, which are prone to runtime failures~\cite{nanz2015study}.

% Problem

Software Fault Injection (SFI) has become a critical technique for evaluating system behavior under such fault scenarios~\cite{ming2020fuzzingerror,chen2020tensorfi,cotroneo2024neural, natella2016assessing}. By injecting controlled \textit{faults} (i.e., \textit{software bugs}), SFI helps identify hidden vulnerabilities and improves system robustness, making it a cornerstone of resilient software design.
A critical challenge in SFI lies in addressing the complexity of \textit{residual faults}, i.e., those hidden defects that elude traditional testing and persist in deployed systems. Achieving fault representativeness is key, as an ideal \textit{faultload} (i.e., the set of faults to inject in the system) should emulate these elusive defects with high fidelity. However, accurately modeling residual faults remains daunting, as it requires replicating the diverse and often unpredictable human error patterns that give rise to software defects. 

While significant research efforts have aimed at identifying and characterizing standard bug classes to approximate residual fault patterns~\cite{titcheu2020selecting,natella2012fault, duraes2006Emulation}, current approaches face three pivotal challenges. First, predefined fault models, often applied through pattern-matching techniques, struggle to capture the complexity and variability of real-world failure scenarios, leading to fault loads that inadequately reflect residual faults~\cite{COSTA2015practicalfaultloads, cotroneo2020profipy}. Then, fault injectors and models are frequently tailored to specific programming paradigms or architectural styles, reducing their applicability across diverse software systems and diminishing their utility for comprehensive fault analysis~\cite{cotroneo2019howbad, cotroneo2021enhancing, cotroneo2022fianalytics, chen2020tensorfi}. Finally, the complexity of configuring and using fault injection tools, along with steep learning curves, discourage widespread adoption and limits their effectiveness against residual faults~\cite{cotroneo2020profipy,huang2012taxonomy,huang2017human,cotroneo2022thorfi}.

% Solution
To address these challenges and advance research on residual faults, we propose \datasetname{}, a comprehensive dataset containing \datasetnumberofbugs{} residual bugs comprising pairs of fault-free and faulty code from major Python open-source frameworks.
The standard driving idea is that faults disclosed by the test cases typically do not represent accurate residual faults \cite{natella2012fault,duraes2006Emulation}, as developers would identify and correct these during the development process. We collected faults by analyzing actual software bugs found in the field, using two sources: GitHub to collect a set of hard-to-find bugs (like concurrency, memory corruption, and security issues) in a custom dataset and established datasets from major conferences and journals. Each fault in our collection links to a specific commit where developers fixed a bug that users found after the software was released - meaning these are bugs that testing didn't catch.

To simplify fault generation for testers and developers who want to assess the behavior of systems against unforeseen faults, we annotated the faults with natural language (NL) descriptions of different levels of detail. This dataset enables the specialization of AI-powered solutions (e.g., Large Language Models, LLMs) to understand fault descriptions and automatically generate corresponding code defects. Users can interact with these models using NL instead of dealing with complex fault models, making fault injection accessible without requiring specialized expertise. Therefore, with \datasetname{}, we aim to bridge the gap between the usage and accessibility of SFI tools while maintaining the complexity of real-world failures by exploiting the potential of NL annotations, a technique widely accepted in the software testing landscape~\cite{nielsen2020use}.

The rest of the paper is structured as follows: 
Section~\ref{sec:related} discusses the related work; 
Section~\ref{sec:dataset} details the process of construction of the dataset and provides statistics; 
%Section~\ref{sec:statistics} provides the statistics of the dataset; 
Section~\ref{sec:conclusion} concludes the paper.

\section{Related Work}
\label{sec:related}
Various datasets have emerged in the domain of Python complex systems, each addressing specific challenges and research goals.
Table~\ref{tab:selected_datasets} summarizes the Python-focused datasets most relevant to our study. These datasets span a wide range of applications, from defect detection and debugging to vulnerability identification and patch analysis. 
%For example, \textit{BugsInPy}~\cite{widyasari2020BugsInPydataset} offers a curated collection of 493 reproducible Python bugs from open-source projects, emphasizing reproducibility for debugging tasks. Similarly, \textit{Defectors}~\cite{mahbub2023defectorsdataset} compiles over 213,000 Python files with labeled instances, making it a valuable resource for defect prediction. In the realm of security, \textit{PySecDB}~\cite{sun2023PySecDBdataset} and \textit{MoreFixes}~\cite{akhoundali2024MoreFixesdataset} focus on Python-specific vulnerabilities and CVE fixes, improving the efficiency and accuracy of vulnerability detection and patch analysis workflows.

Despite their contributions, existing datasets often lack certain features essential for fault injection studies. Most datasets are designed to address general-purpose tasks, such as bug detection or patch analysis, rather than focusing on the challenges posed by residual faults. While less frequently studied, these faults are critical for understanding complex failure modes in real-world software systems.
Another limitation of existing datasets is the granularity of metadata, as they lack detailed descriptions of fault behavior, which are crucial for advanced fault analysis and NL-driven fault injection. In contrast, our dataset enriches fault instances with NL descriptions, enabling more effective integration with AI-driven fault analysis tools.
Furthermore, prior datasets often target narrowly defined tasks, such as vulnerability detection or framework-specific bug analysis, without offering a comprehensive view of functional fault diversity. 
%For instance, \textit{FrameworkFaults}~\cite{du2023frameworkfaulttriggersdataset} examines bugs in software frameworks, while \textit{SilentBugs}~\cite{Tambon2024silentbugstensorflowdataset} focuses on undetected issues in TensorFlow. 
%These datasets provide valuable insights within their respective domains but do not generalize well across diverse Python projects.

Our dataset addresses these gaps by combining the strengths of existing resources while introducing unique features tailored for fault injection studies. It aggregates diverse residual faults from multiple Python projects, ensuring functional diversity and reproducibility. Moreover, including detailed metadata supports advanced fault analysis, bridging the gap between reproducibility and AI-enabled evaluation. These features position our dataset as a valuable tool for researchers exploring fault injection, debugging, and security in Python systems.

\section{\datasetname{} Dataset}
\label{sec:dataset}
This section details the creation of \datasetname{}, a dataset comprising \datasetnumberofbugs{} residual faults. These faults escaped detection during testing and release, only to surface later in operation, as evidenced by bug reports. Their occurrence highlights test suite limitations and the need for enhanced fault detection mechanisms \cite{natella2012fault, duraes2006Emulation, xuemei2003faultremoval}.

The rest of the section details the steps involved in creating and curating the dataset. We begin by describing the process of fault collection (\S{}~\ref{subsec:data_collection}), focusing on identifying and sourcing residual bugs from real-world projects. Next, we elaborate on the filtering and selection criteria applied to ensure the dataset's quality and relevance (\S{}~\ref{subsec:data_filtering}). Finally, we explain how we documented the faults using multi-level natural language descriptions (\S{}~\ref{subsec:fault_description}) and systematically categorized (\S{}~\ref{subsec:statistics}), enabling their use in AI-driven fault injection and analysis.

\begin{table}[t]
\caption{List of Selected Python Datasets.}
\label{tab:selected_datasets}
\scriptsize
\begin{tabular}{
>{\arraybackslash}m{1.5cm} |
>{\arraybackslash}m{5cm} |
>{\arraybackslash}m{1cm}
}
\toprule
\textbf{Author(s) (Year)} & \textbf{Description and Contribution} & \textbf{Focus}\\
\midrule
Cotroneo \textit{et al.} (2019)~\cite{cotroneo2019howbad} & OpenStack bug study: 179 critical fail-stop faults analyzed. & Fault Analysis \\
\midrule
Widyasari \textit{et al.} (2020)~\cite{widyasari2020BugsInPydataset} & BugsInPy: 493 reproducible bugs from Python projects, aiding testing and debugging. & Bug ~\mbox{Benchmarking} \\
\midrule
Akimova \textit{et al.} (2021)~\cite{akimova2020pytracebugs} & PyTraceBugs: 24k buggy and 5.7M correct code samples from 11k+ repos. & DL Training \\
\midrule
Bhandari \textit{et al.} (2021)~\cite{bhandari2021CVEFixesdataset} & CVEFixes: Links 5,365 CVE records to 5,495 vulnerability fixing commits from 1,754 projects. & Vulnerability Analysis \\
\midrule
Xu \textit{et al.} (2022)~\cite{xu2022patchtracerdataset} & Tracer: A framework for tracking vulnerability patches with a dataset of 1,295 CVEs. & Vulnerability Analysis \\
\midrule
Sun \textit{et al.} (2022)~\cite{sun2023PySecDBdataset} & PySecDB: Python-focused dataset of 1,258 security commits and 2,791
 non-security commits. & Vulnerability Analysis \\
\midrule
Mahbub \textit{et al.} (2023)~\cite{mahbub2023defectorsdataset} & Defectors: Includes 213K Python files with 93K labeled instances for defect prediction. & DL Training \\
\midrule
Richter \textit{et al.} (2023)~\cite{richter2023nbfdataset} & A 33k Python bug fix dataset achieving 170\% better neural bug detection vs artificial data. & DL Training \\
\midrule
Akhoundali \textit{et al.} (2023)~\cite{akhoundali2024MoreFixesdataset} & MoreFixes: Curates 26,617 validated CVE fixes for security analysis. & Vulnerability Analysis \\
\midrule
Du \textit{et al.} (2023)~\cite{du2023frameworkfaulttriggersdataset} & Analyzes 3,555 bugs from DL-frameworks, with a focus on their classification and characteristics. & Fault Analysis \\
\midrule
Tambon \textit{et al.} (2023)~\cite{Tambon2024silentbugstensorflowdataset} & SilentBugs: Analysis and classification of 77 "silent bugs" TensorFlow faults. & Fault Analysis \\
\bottomrule
\end{tabular}

\vspace{-0.3cm}
\end{table}

\begin{figure*}[t]

\centering
\noindent
{\footnotesize python/cpython commit 3b959dbcaf}

\begin{minipage}{0.6\columnwidth}
\centering
{\footnotesize Fault Free Code}
\begin{lstlisting}[style=pythonstyle]
def put(self, item, block=1):
    # Same as before self.mutex.acquire()
    (*@\colorbox{codeadd}{release\_fsema = True}@*) 
    try:
        (*@\colorbox{codeadd}{was\_empty = self.\_empty()}@*)
        self._put(item)
        if was_empty:
            self.esema.release()
        release_fsema = not self._full()
    finally:
        if release_fsema:
            self.fsema.release()
        self.mutex.release()
\end{lstlisting}
\end{minipage}
\hspace{0.1\textwidth}  
\begin{minipage}{0.6\columnwidth}
\centering
{\footnotesize Faulty Code}
\begin{lstlisting}[style=pythonstyle]
def put(self, item, block=1):
    if block:
        self.fsema.acquire()
    elif not self.fsema.acquire(0):
        raise Full
    self.mutex.acquire()
    was_empty = self._empty()
    self._put(item)
    if was_empty:
        self.esema.release()
    (*@\colorbox{coderemove}{if not self.\_full():}@*)
        self.fsema.release()
    self.mutex.release()
\end{lstlisting}
\end{minipage}

\vspace{1em}  % Spazio tra il codice e l'immagine
\includegraphics[width=1\textwidth]{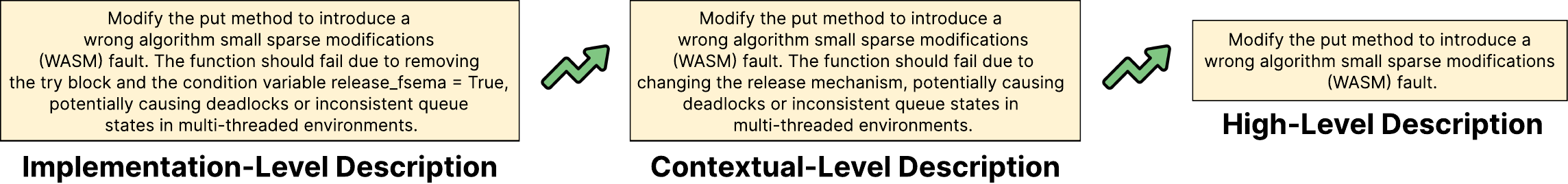}

% Caption e label generali
\caption{Example of a deadlock fault in CPython: faulty and fault-free code versions, with three NL descriptions for bug injection in the \textit{put} method}
\label{fig:multilevel_descriptions}

\vspace{-0.5cm}
\end{figure*}

\subsection{Data Collection}
\label{subsec:data_collection}

To collect faults and build the \datasetname{}, we systematically selected corpora from top-tier conferences and journals in the software engineering field, focusing on research published in the last 5 years. 
A critical insight guiding our approach, based on prior studies \cite{fonseca2008mapping, fonseca2014analysis, cerveira2017vulnerabilityemulation}, is the recognition that vulnerabilities often represent residual faults—those escaping test suites and manifesting later as zero-day exploits. This assumption led us to incorporate Common Vulnerabilities and Exposures (CVEs) into the dataset, further enriching its utility for fault injection and defect prediction research.

Overall, we selected $11$ datasets that span diverse domains, including software fault injection, automatic program repair, defect prediction, controlled testing, and vulnerability analysis,  described in \tablename{}~\ref{tab:selected_datasets} 

To further enhance the dataset’s relevance and utility, we enriched it with additional residual fault categories critical for system testing yet underrepresented in existing datasets. These fault types, often particularly challenging for developers to identify, debug, and fix~\cite{cotroneo2013faultriggers, piantadosi2019fixing}, include Mandelbugs (e.g., aging-related bugs), Heisenbugs (e.g., concurrency issues), memory corruption bugs, high-priority or critical bugs, and security vulnerabilities. These categories were selected due to their significant impact on software reliability and the inherent difficulty in addressing them, making their inclusion essential for a comprehensive representation of real-world faults.

We carefully selected faults from GitHub repositories to ensure accurate representation by leveraging metadata such as bug reports, commit messages, and linked discussions to validate the fault's nature. This enrichment resulted in 67 additional faults, further broadening the scope and applicability of \datasetname{} for fault injection and automated testing research.

To standardize data, we designed a composite key with commit hash, project URL, faulty code, and fixed code, ensuring traceability, reproducibility, and seamless analysis.

\subsection{Data Filtering}
\label{subsec:data_filtering}
%On the faults collected during the data collection phase, we performed data filtering to focus on meaningful and non-trivial faults.

We applied a rigorous filtering process to ensure that \datasetname{} meets its primary objective of representing residual bugs. We guided this process by two core requirements, specifically designed to identify and include faults that align with the definition of residual bugs:

\begin{enumerate}
    \item The fault must originate from the source code, specifically within a method implementation, and exclude issues related to configuration files, documentation, or build scripts to include only actual bugs encountered during software development.
    %This guarantees empirical validity by including only actual bugs encountered during software development, ensuring the dataset focuses on faults with practical relevance.

    \item The fault must have a corresponding bug report (identified via commit hashes, pull requests, issues, or formal bug reports) documenting its faulty and faulty-free version (i.e., the bug fix). This criterion ensures the representativeness of residual faults and supports reproducibility by focusing on verifiable defects encountered in real-world scenarios.
\end{enumerate}

We focused on faults from widely adopted, complex Python projects to ensure dataset quality and relevance. We selected projects from diverse domains, including prominent frameworks and libraries such as pandas, CPython, Django, Ansible, Apache Airflow, scikit-learn, NumPy, Black, OpenStack, and Scrapy. We based project selection on community adoption metrics, with repository star counts ranging from 2.5k (Pycripto) to 187k (TensorFlow). While we primarily included faults from these high-impact projects, we made strategic exceptions for critical security vulnerabilities from less popular repositories to ensure comprehensive coverage of representative fault examples.

%Then, we extracted the faulty code and its corresponding patch for each fault, maintaining the code at the method level \hl{due to technological limits} \cite{Add citation for technological limits} \giuseppe{Aggiunto per rispondere alla domanda: perché method-level?}. \hl{Although we acknowledge that many residual faults stem from configuration errors and non-code-related issues, we focus on code generation tasks, specifically in Python, to support model training for fault generation}. \giuseppe{Modificato per rispondere alla domanda: perché solo fault del codice?} We employed specialized GitHub scraping tools to facilitate this process. To enhance clarity and reproducibility, we also extracted additional metadata, including commit messages, bug type classifications, CVE IDs (if available), associated test cases, and the Python version used.

Then, we extracted each faulty code snippet and its corresponding patch at the method level. Our goal was to capture self-contained bugs that can affect multiple parts of the program through a single function or method call, making them particularly relevant for software fault injection. Although we acknowledge that many residual faults may arise from configuration errors or issues outside the code, we opted to focus on Python source code to train generative models tailored to code-centric fault injection. Consequently, we excluded non-code artifacts such as configuration files to consistently emphasize code-level defects. We employed specialized GitHub scraping tools to automate data collection, and to enhance clarity and reproducibility, we also extracted metadata such as commit messages, bug type classifications, CVE IDs (if available), associated test cases, and the Python version used.

Finally, we deduplicated using \texttt{[faulty code, fixed code]}, filtered invalid entries, and excluded comment-only changes to ensure consistency and remove redundancy. This process retained \datasetnumberofbugs{} fault pairs.

\subsection{Fault Description}
\label{subsec:fault_description}
We developed a comprehensive approach to describing faults to support the generation of faults directly from NL descriptions, hence streamlining the SFI process. To help model training across various prompt formats and simulate human tester behavior, we generated three levels of NL descriptions for each fault: \textit{Implementation-Level Description}, \textit{Contextual-Level Description}, and \textit{High-Level Description}.

To understand the differences in the three NL descriptions, \figureautorefname{}~\ref{fig:multilevel_descriptions} illustrates the progression of fault descriptions for a bug in CPython's \texttt{put} method, showcasing the three levels of natural language. The figure visually ties these descriptions to the associated code changes in the Faulty Code (right) and Fault-Free Code (left), demonstrating how each level provides progressively less technical detail while retaining relevance for different fault modeling and testing purposes.

The Implementation-Level Description (bottom left) provides the most detailed and technical view, specifying the code changes needed to introduce the fault. It highlights the removal of the try block and the condition \texttt{release\_fsma = True}, explaining how these changes cause deadlocks or inconsistent queue states in a multi-threaded environment. 

Instead, the Contextual-Level Description (center bottom) slightly abstracts the fault by focusing on its general mechanism and potential impact rather than specific code modifications. It mentions modifying the put method and altering the release mechanism, leading to potential issues such as deadlocks or inconsistent states but avoids specifying exact code lines. This level provides testers with a broader understanding of the fault's behavior and consequences.

In the High-Level Description (bottom right), we make the description entirely abstract and omit technical or contextual details about the specific fault. Modifying the put method introduces a ``\textit{wrong algorithm small sparse modifications fault}" in the fault-free function. This description suits scenarios where a conceptual understanding of the fault type is sufficient without providing implementation specifics.

%The fault descriptions were created and validated by a team of six researchers with expertise in computer engineering and cybersecurity, including one postdoctoral researcher with a PhD in information technologies and a strong background in AI and fault injection, one PhD student specializing in cybersecurity, and four M.Sc. thesis students in computer engineering. Moreover, a full professor in computer engineering with extensive experience in software testing and fault injection supervised the process, \hl{reviewing and approving a subset of the final descriptions} \giuseppe{Da capire come porla}.
%Over six months, the team generated and validated descriptions for all the faults in the dataset. \hl{The PhD student finally reviewed all the descriptions.} \giuseppe{Può aver senso?}
%This process also served to identify and filter out fault pairs that involved non-representative fault classes.

A team of six researchers specialized in computer engineering and cybersecurity created and validated the fault descriptions, under the coordination of a full professor with extensive expertise in software testing and fault injection. The professor established the description style, while the postdoctoral researcher, with a PhD in information technologies and background in AI and fault injection, provided ongoing reviews and feedback. The team, which also included a PhD student in cybersecurity and four M.Sc. thesis students, worked together over six months to generate and validate the descriptions. Finally, the professor, postdoc, and PhD student conducted a comprehensive review to ensure consistency and quality.

We involved multiple researchers in generating fault descriptions to ensure quality, reduce workload, and maintain accuracy. We preferred human-generated descriptions over AI-based solutions to avoid biases and reliance on repetitive phrasing. Moreover, the team's diverse linguistic styles and technical expertise introduced natural variability, enhancing the dataset's robustness and improving the generalization capabilities of downstream models.

The final dataset integrates the \datasetnumberofbugs{} collected faults with their corresponding three levels of NL descriptions. Each sample in \datasetname{} includes the three NL descriptions (Implementation-Level, Contextual-Level, and High-Level) detailing how to inject the fault, the corresponding fault-free code (the base for injecting the fault), and the resulting faulty code generated by applying the described fault.

We analyzed the complexity of tokens, the NL descriptions, and both the faulty and fault-free functions.
The NL descriptions accompanying the faults add an extra layer of richness to the dataset. High-level descriptions, which are the most abstract, average $14$ tokens (SD: $3.3$9), while Contextual-Level Descriptions average 32 tokens (SD: $5.6$). Implementation-level descriptions, the most detailed, average $34$ tokens (SD:  $4.8$). This progression reflects the growing detail and technicality of the descriptions, meeting diverse research needs.

The analysis of fault-free code and its faulty code counterparts shows minimal variation in token counts. Fault-free code snippets average $74$ tokens (SD: 104), while Faulty Code snippets average $71$ (SD: 103). This balance highlights that the faults are typically well-localized and do not result in excessive or extraneous code changes, maintaining the dataset’s focus on precise and meaningful fault examples.

% Table of bug distribution in our custom dataset
%\begin{table}[ht]
%\centering
%\begin{tabular}{|l|c|c|}
%\hline
%\textbf{Bug Category} & \textbf{Occurrences} & %\textbf{Percentage} \\
%\hline
%Security & 446 & 92.7\% \\
%Mandelbugs & 15 & 3.1\% \\
%Mem Corruption & 15 & 3.1\% \\
%Heisenbugs & 6 & 1.1\% \\
%\hline
%\textbf{Total} & 482 & 100\% \\
%\hline
%\end{tabular}
%\caption{Distribution of Bug Types of the custom dataset}
%\label{tab:custom_dataset_bug_distribution}
%end{table}

\subsection{Fault Classification}
\label{subsec:statistics}
%Our analysis revealed a comprehensive fault distribution over the software systems integrated. \figureautorefname{}~\ref{fig:fault_distribution} shows the top 10 of the fault types that dominate our dataset, in which the top three are \textit{Missing IF construct plus Statements} (MIFS): 10.08\%, \textit{Missing Parameter in Function Call} (MPFC): 10.41\%, and \textit{Wrong variable used in Parameter of Function Call} (WPFV): 5.39\%.

\begin{comment}
\begin{enumerate}
\item \textit{Missing IF construct plus Statements} (MIFS): 10.08\% 
\item \textit{Missing Parameter in Function Call} (MPFC): 10.41\% 
\item \textit{Wrong variable used in Parameter of Function Call} (WPFV): 5.39\% 
\end{enumerate}
\end{comment}

To systematically categorize these residual faults, we adopted the extended version of the Orthogonal Defect Classification (ODC) framework developed by Duraes~\textit{et al.}~\cite{duraes2006Emulation}, with further refinements from subsequent studies~\cite{fonseca2008mapping, pereira2016practical}. This classification schema, tailored for SFI, is designed to identify and emulate realistic software faults. It categorizes defects based on their characteristics (e.g., whether the code is missing, incorrect, or extraneous) and the nature of the fault (e.g., assignments, conditionals, function calls, or algorithms), containing a total of $61$ different fault categories. 

\begin{comment}
\begin{figure}[t]
\centering
\includegraphics[width=0.6\columnwidth]{images/top10_pie_faults.pdf}
\vspace{-0.2cm}
\caption{Distribution of Fault Types in the Dataset.}
\vspace{-0.5cm}
\label{fig:fault_distribution}
\end{figure}
\end{comment}

To comprehensively understand the dataset, we analyzed the distribution of fault types across the integrated software systems. The dataset encompasses $56$ distinct fault types out of a possible $61$ categories, with the remaining five categories absent due to their association with C-style programming patterns, which are uncommon in Python. %This broad coverage underscores the dataset's focus on capturing Python-specific fault patterns, reflecting real-world issues encountered in dynamically typed environments.

We also analyzed the most frequent fault categories to highlight prevalent challenges in Python software development (we limited the discussion to the top 3 due to space limit). Three key fault types dominate the dataset: Missing Parameter in Function Call (MPFC), accounting for $10.78\%$ of the dataset; Missing IF Construct plus Statements (MIFS), contributing $10.11\%$; and Wrong Algorithm - Large Modifications (WALL), representing $7.92\%$.
According to the ODC, these categories address critical aspects of software faults. MPFC faults, categorized as Interface issues, arise when required parameters are omitted in function calls, often causing runtime errors. MIFS faults, under Checking, involve the absence of conditional logic (e.g., if statements), leading to incorrect execution paths. WALL faults, classified as Algorithm issues, represent significant errors in algorithm implementation, such as flawed logic or design structure. Other fault categories in our dataset include Wrong Function Called with Different Parameters (WFCD, $6.64\%$), Wrong Variable Used in Parameter of Function Call (WPFV, $5.77$\%) and Wrong Function Called with Same Parameters (WFCS, $5.57$\%).

%\section{Dataset Statistics}
%\label{sec:statistics}
%\input{dataset_statistics/statistics}

\section{Conclusion and Future Work}
\label{sec:conclusion}
We introduced \datasetname{}, the first dataset specifically designed for SFI in Python applications, focusing exclusively on residual faults. The dataset comprises \datasetnumberofbugs{} curated pairs of faulty and fault-free code snippets extracted from real-world Python projects, each enriched with three levels of NL descriptions and detailed metadata.

The dataset is publicly available on GitHub~\footnote{ \url{https://github.com/dessertlab/PyResBugs}}. Future work will focus on developing an SFI tool powered by a model trained on \datasetname{}, showcasing the dataset’s effectiveness and advancing the field of automated fault injection.

\section{Acknowledgments}
\label{sec:ack}
Work supported by the \textit{IDA—Information Disorder Awareness} Project, funded by the EU–Next Generation EU under the SERICS program (Grant PE00000014, MUR NRRP).

%the \textit{IDA—Information Disorder Awareness} Project funded by the European Union-Next Generation EU within the SERICS Program through the MUR National Recovery and Resilience Plan under Grant PE00000014.

%\section*{Acknowledgment}
%Omitted for double-blind review.

% Per ora utile a dividere le reference dal testo

\bibliographystyle{ieeetr}
\small{
    \bibliography{biblio}
}

\end{document}